\documentclass[12pt]{article}
\usepackage{graphics}
\begin{document}
\begin{flushright}
WISC-MILW-98-TH-15 \\
{\em Physical Review C, vol. 57, pg.1525-1527 (1998)}\\
\end{flushright}
\vspace{.4in}
\begin{center}
{\Huge Heavy ion beam lifetimes at relativistic} \vspace{3mm}\\
{\Huge and ultrarelativistic colliders}\\
\vspace{.3in}
{\Large John W. Norbury}\\
\vspace{.15in}
{\em Physics Department, University of Wisconsin - Milwaukee,}\\
{\em P.O. Box 413, Milwaukee, Wisconsin 53201.}\\ (email: norbury@uwm.edu)\\
\vspace{.25in}
{\Large Marsha L. Waldsmith}\\
\vspace{.15in}
{\em Physics Department, University of Wisconsin - La Crosse,}\\
{\em La Crosse, Wisconsin 54601.} \\
\end{center}
\vspace{5mm} \noindent

\begin{center}
{\bf Abstract}
\end{center}
\noindent
The effects of higher order corrections in ultra-relativistic nuclear collisions are considered. It is found that
higher order contributions are small at low energy, large at intermediate energy and small again at very high
energy. An explanation for this effect is given. This means that the Weizsacker-Williams formula is a good
approximation to use in calculating cross sections and beam lifetimes at energies relevant to RHIC and LHC.

\vspace{10mm}

\newpage
Heavy ion and relativistic heavy ion accelerators have so far been constructed with an incident beam and a
fixed target. The first heavy ion {\em collider}, named the Relativistic Heavy Ion Collider (RHIC)
\cite{rhic}, is currently under construction at Brookhaven National Laboratory. This machine will collide
heavy ions all the way from the proton up to Pb nuclei with an energy per beam of 100 GeV/nucleon. This is
equivalent to a fixed target accelerator with a beam energy of 21.7 TeV/nucleon. The Large
Hadron Collider (LHC) is currently being planned for the purposes of proton-proton collisions. However LHC
will also be capable of accelerating heavy ions. One of the main purposes in constructing these two
relativistic heavy ion colliders is to undertake a search for the quark gluon plasma phase.

Heavy nuclear beams installed in heavy ion colliders will only have a finite lifetime due mainly to  beam-beam
Coulomb dissociation and  electron pair production.  In designing heavy ion colliders it is
important to have good estimates for the various beam lifetimes \cite{rhic}. Such an estimate for U-U beams
installed in RHIC have been carried out
\cite{Norbury}. However in that work the effects of higher order electromagnetic processes
\cite{explanation} was left out. Such effects are known to be small (about 4\%) at low energy
\cite{Wheeler}. However these effects have been found to be larger (about 12\%) at higher energies
\cite{explanation}. Thus the question immediately arises as to how big these effects might become at the very high
energies planned for RHIC and LHC. In particular, if the effects are quite large then previous estimates of the
beam lifetimes \cite{rhic, Norbury} might be incorrect. This may  have a significant effect on accelerator
operations.

Generally the higher order cross sections are smaller than the Weizsacker-Williams cross section. In multiple
electromagnetic  processes the excitation of the single photon giant dipole resonance is depleted by multiple
higher order electromagnetic processes, which can excite double and multiple giant resonances, generally lying at
double and multiple integral values of the energy of the single dipole state. The  Weizsacker-Williams 
calculation assumes that all of the photons excite only the single giant dipole resonance state.

Calculations were performed for the Coulomb dissociation cross sections, $\sigma_{WW}$, using the
Weizsacker-Williams method \cite{BB} for Uranium-Uranium (UU) collisions at a variety of energies. The calculations
were repeated for the cross sections including higher order corrections, $\sigma_N$,  as discussed in
\cite{explanation}. The percentage difference,  $100(\sigma_{WW}-\sigma_N)/\sigma_{WW}$, is then
calculated. The results for fission of one of the U nuclei are presented in Figure 1. As discussed above, it is
seen that the higher order effects are small at low energy and then grow as the energy is increased. However the
surprising feature is that as the energy is increased even further, the higher order effect starts getting smaller
again.  For RHIC and LHC energies the percentage
difference is less than 2\%. This is good news for calculations of beam lifetimes, because it means that the old
Weizsacker-Williams calculations
\cite{rhic, Norbury} remain correct. It has been found previously \cite{Norbury1} that other corrections, such
as Rutherford bending and electric quadrupole effects, can also be neglected at high energy. Thus at the very high
energies of RHIC and LHC one can essentially ignore all corrections (including higher order corrections), and just
rely on the basic Weizsacker-Williams formula. The calculations represented in Figure 1 are for fission, but similar
results were also found for other Coulomb dissociation reactions such as single and double nucleon removal and
also for other nuclei.

Further analysis revealed the source of the behavior seen in Figure 1. In performing a Weizsacker-Williams
calculation one folds and integrates the virtual photon spectrum with the photonuclear cross section as in
\begin{equation}
\sigma_{WW} = \int N(E) \sigma (E) dE
\label{1}
\end{equation}
where $\sigma$ is the nucleus-nucleus Coulomb dissociation cross section, $N(E)$ is the virtual photon spectrum as
a function of virtual photon energy $E$, and $\sigma(E)$ is the photonuclear cross section. The
virtual photon spectrum is a smoothly dropping function of virtual photon energy \cite{BB}, while the photonuclear
cross section is a bell shaped curve centered on a fixed photon energy. As the heavy ion energy is increased the
amount of ``overlap" between the virtual photon spectrum and the photonuclear cross section  steadily increases.
In impact parameter space this cross section is written \cite{explanation}
\begin{equation}
\sigma = 2\pi \int P(b) bdb
\label{2}
\end{equation}
where $P(b) = PWW(b)$ is the Weizsacker-Williams probability given by
\begin{equation}
PWW(b) = \Phi (b)
\label{3}
\end{equation}
where
\begin{equation}
\Phi (b) =  \int n(E,b) \sigma (E) dE/E
\label{4}
\end{equation}
with $n(E,b)$ being the impact parameter dependent virtual photon spectrum \cite{explanation}.
Thus if $PWW(b)$ is used for $P(b)$ in equation (\ref{2}) then one reproduces the numerical values obtained from
(\ref{1}). The higher order cross section is calculated with equation (\ref{2}) except that
\begin{equation}
PN(b) = \Phi (b) e^{-\Phi (b)}
\label{5}
\end{equation}
is used for the probability $P(b)$. This Poisson probability distribution results from the statistical independence
of the emission of a number $N$ of photons emitted from a classical current source \cite{IZ}. 

Now to explain Figure 1.  The probabilities, multiplied by $b$, (since
it is $b$ x {\em probability} that gets integrated in (\ref{2})) are plotted in Figures 2 - 4.
At very low energies $\Phi(b)$ will be very small, so that $\Phi (b) $ and
$\Phi (b) e^{-\Phi (b)}$ will be approximately the same leading to equality of the  Weizsacker-Williams cross
section $\sigma_{WW}$ and the higher order cross section $\sigma_N$. This is seen in Figure 2, and is reflected in
Figure 1 at the low energy end where the Weizsacker-Williams and higher order cross sections are approximately the
same. As the energy is increased however, $\Phi(b)$ increases and  $\Phi (b) e^{-\Phi (b)}$ starts getting 
smaller than $\Phi (b)$, as shown in Figure 3, meaning that $\sigma_N$ is smaller than $\sigma_{WW}$ which explains
the rising part of Figure 1 between $10^{-2}$ GeV/nucleon and $10^0$ GeV/nucleon. Remember that it is the areas
under the curves in Figures 2 - 4 that gives the cross sections according to (\ref{2}).

Figure 4 shows the probabilities (times $b$) at very high projectile energy. Even though there is still a
difference between the probability curves, however now both curves extend out to much larger impact paramteres
where the differences are much smaller. (High energies probe larger impact parameters.) This results in the total
areas under the two curves in Figure 4 being much closer in value to each other than the two curves in Figure 3.
(This was seen by explicit computation.) This means that for the case of Figure 4 the resulting cross sections
(calculated from (\ref{2})) are in fact quite close to each other. This explains the small percentage difference
value in Figure 1 at high energy. Consequently, the curve in Figure 1 has reached some maximum value (near 1
GeV/nucleon) and then starts falling as the energy increases. These various effects then explain the behaviour seen
in Figure 1.

At very high energy then, one can safely use just the Weizsacker-Williams formula for calculating beam lifetimes,
even for the heaviest of nuclei where one might have thought that higher order effects might have been important.
As an application of this a Uranium beam lifetime is calculated for LHC. The method of calculation is the same as
described previously \cite{Norbury}. The beam parameters \cite{rpp} are listed in Table 1, where $k$ is the number
of beam intersections,
$L_0$ is the initial luminosity, $B$ is the number of bunches and $N_B$ is the number of particles per bunch. Cross
sections and lifetime results are listed in Table 2, where some previous results \cite{Norbury, Baur} for RHIC are
also listed for comparison. (The greatest uncertainty in these calculations is the pair production cross section
$\sigma_{e^+ e^-}$ \cite{Baur}. Also in \cite{Norbury}, the value for $\sigma_{e^+ e^-}$ should have been 85 barn.)
Table 2 shows that Uranium beams will live sufficiently long so that they could be installed at RHIC and LHC.

In conclusion, it has been shown that higher order contributions to Coulomb dissociation are small at low energy,
large at intermediate energy and then small again at very high energy. Consequently, the use of the
Weizsacker-Williams formula in calculating cross sections and beam lifetimes is expected to be accurate at RHIC
and LHC energies.

This work was supported by NASA grant 97-367-NM.

\newpage

\begin{tabular}{|c|c|c|} \hline
Beam parameter & RHIC & LHC \\ \hline
k &6 & 1  \\
$L_0$ ($cm^{-2}$ $s^{-1}$) & $9.2 \times 10^{26}$  & $2 \times 10^{27}$ \\
B & 57 & 608 \\
$N_B$ & $1.1 \times 10^9$ &  $9.4 \times 10^7$ \\
energy/nucleon/beam (GeV) & 100 & 2760 \\  \hline
\end{tabular}

\vspace{10mm}
\hspace{1in}Table 1: Beam Parameters\\

\vspace{2in}

\begin{tabular}{|c|c|c|c|c|c|} \hline
 &  & $\sigma$ (barns) & & &  $\tau$ (hours)\\ 
 &fission & 1N   & 2N  & $e^+ e^-$ & \\ \hline
RHIC & 33 & 48  & 31 & 85  & 11\\
LHC & 56 & 80 & 52 & 135 &  17 \\ \hline
\end{tabular}

\vspace{10mm}
\hspace{.6in}Table 2: Weizsacker-Williams Results\\

\newpage

\includegraphics{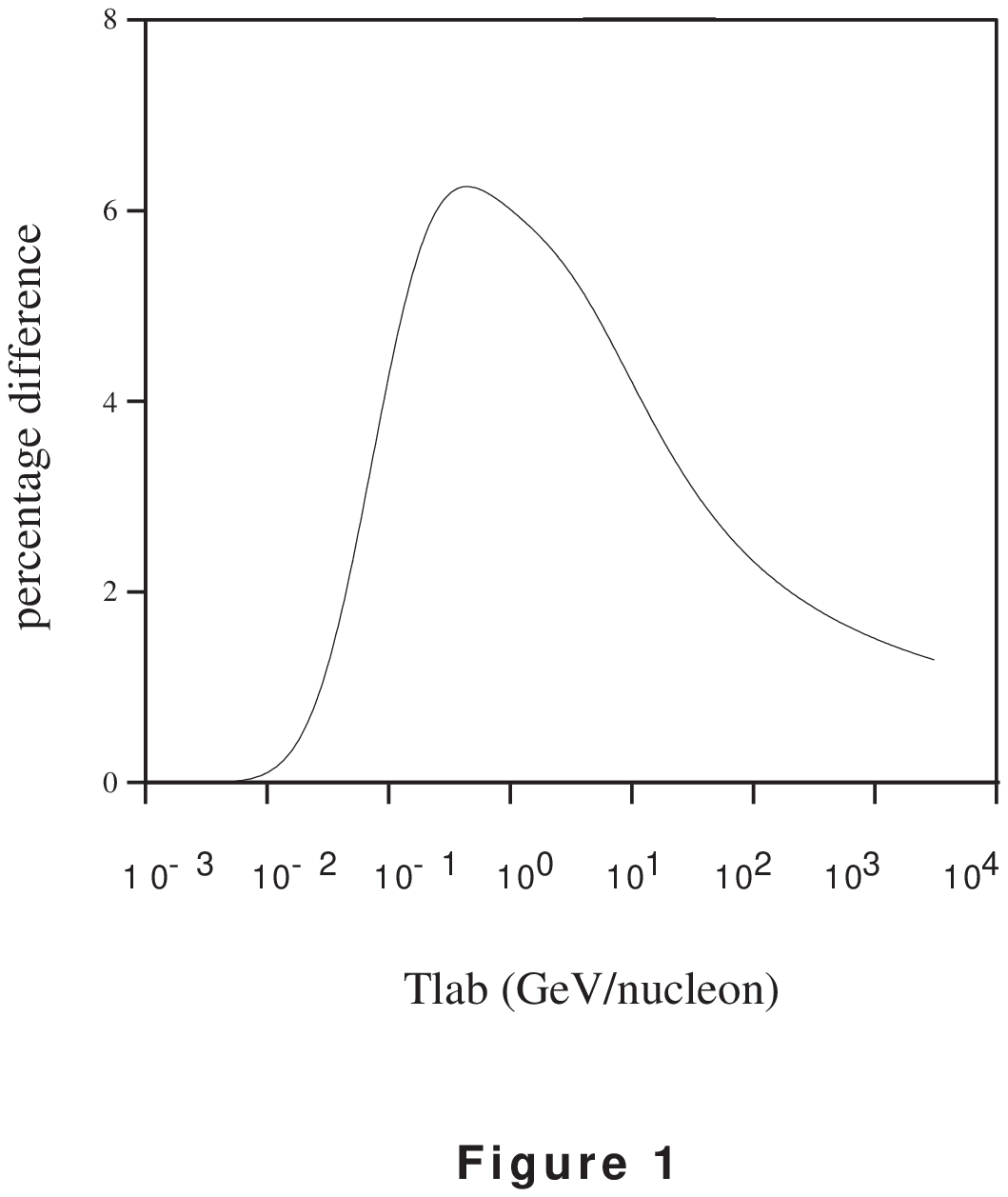}

\vspace{2mm} \noindent
Percentage difference between Weizsacker-Williams and higher order cross sections for 
fission in UU collisions as a
function of projectile kinetic energy for an equivalent fixed target accelerator.
The RHIC energy (21,700 GeV/nucleon) and the LHC energy (16,237 TeV/nucleon) lie  beyond the far right
of the graph.

\includegraphics{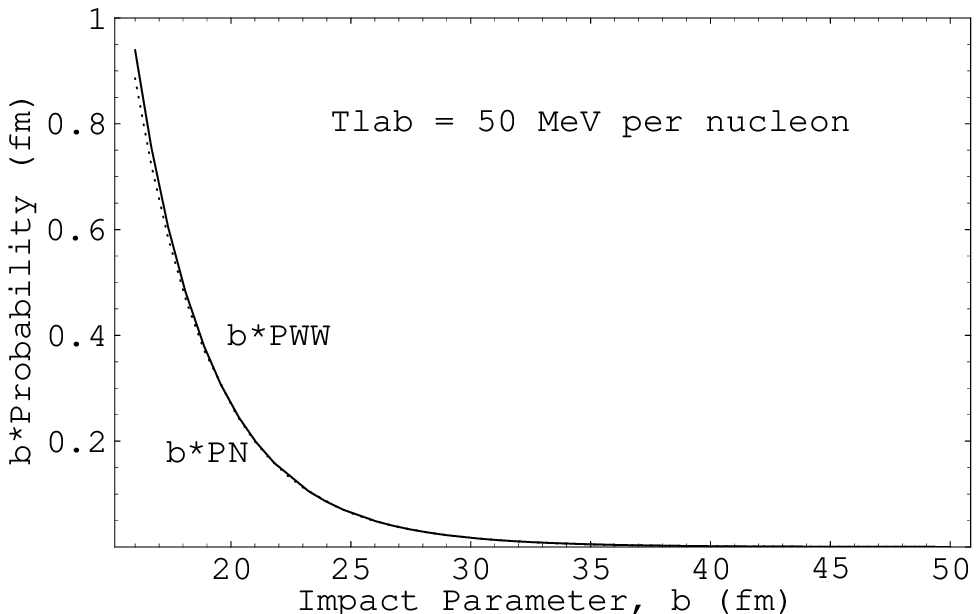}

\vspace{10mm} \noindent

\begin{center}
{\bf Figure  2}
\end{center}
\begin{tabbing}
xxx\=xxxxxxxxxxx\kill
\>Impact parameter multiplied by probability as a function of impact parameter\\ 
\>for Weizsacker-Williams and higher order theory for a projectile energy of \\
\>50 MeV/nucleon. The area
under each curve gives the total cross section.

\end{tabbing}

\newpage
\includegraphics{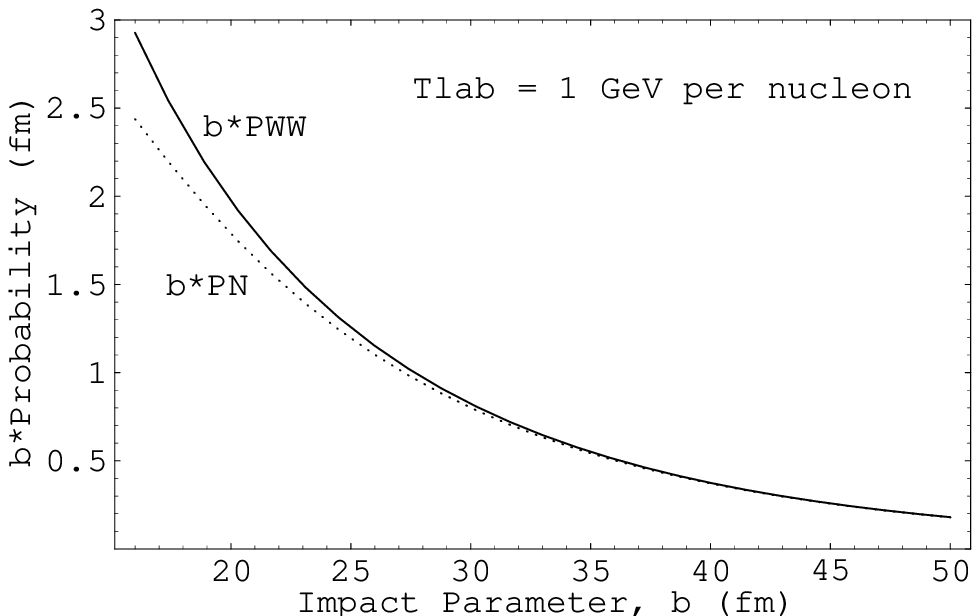}

\vspace{10mm} \noindent
\begin{center}
{\bf Figure  3}
\end{center}
\hspace{.5in} Same as Figure 2 except for a projectile energy of 1 GeV/nucleon.

\newpage
\includegraphics{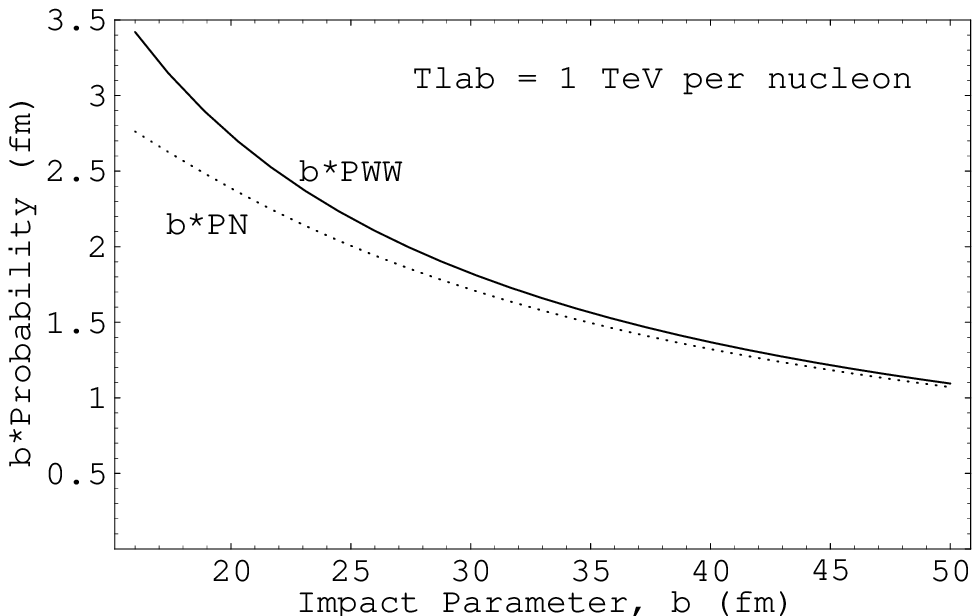}

\vspace{10mm} \noindent
\begin{center}
{\bf Figure  4}
\end{center}
\hspace{.5in} Same as Figure 2 except for a projectile energy of 1 TeV/nucleon.

\end{document}